\documentclass[preprint,notoc]{JHEP3}
\usepackage{epsfig}
\def\eslt{E_T^{\rm miss}}

\def\to{\rightarrow}

\def\bi{\begin{itemize}}
\def\ei{\end{itemize}}

\def\tst{\tilde t}
\def\ttau{\tilde \tau}

\def\tg{\tilde g}

\def\tq{\tilde q}

\def\tw{\widetilde W}
\def\tz{\widetilde Z}
\def\alt{\stackrel{<}{\sim}}
\def\agt{\stackrel{>}{\sim}}
\def\be{\begin{equation}}  
\def\ee{\end{equation}}  
\def\bea{\begin{eqnarray}}  
\def\eea{\end{eqnarray}}  

\title{Implications of Compressed Supersymmetry\\
for Collider and Dark Matter Searches}
\author{Howard Baer$^a$, Andrew Box$^b$, Eun-Kyung Park$^a$, 
and Xerxes Tata$^b$\\
$^a$Department of Physics, Florida State University Tallahassee, 
FL 32306, USA\\
$^b$Department of Physics and Astronomy, University of Hawaii,
Honolulu, HI 96822, USA\\
E-mail: \email{baer@hep.fsu.edu},\email{abox@phys.hawaii.edu},
\email{epark@hep.fsu.edu},\email{tata@phys.hawaii.edu}}

\preprint{\vbox{FSU-HEP-070704, UH-511-1108-07}}

\abstract{
Martin has proposed a scenario dubbed ``compressed supersymmetry''
(SUSY) where
the MSSM is the effective field theory between energy scales
$M_{\rm weak}$ and $M_{\rm GUT}$, but with
the GUT scale $SU(3) $ gaugino mass $M_3\ll M_1$ or $M_2$. As a result, 
squark and gluino masses are suppressed relative to slepton, chargino
and neutralino masses, 
leading to a compressed sparticle mass spectrum, and where
the dark matter relic density in the early universe may be dominantly governed
by neutralino annihilation into $t\bar{t}$ pairs via exchange of a 
light top squark. We explore the 
dark matter and collider signals expected from compressed SUSY
for two distinct model lines with differing assumptions about GUT scale
gaugino mass parameters.
For dark matter signals, the compressed squark 
spectrum leads to an enhancement  in direct detection rates compared 
to models with unified gaugino masses.
Meanwhile,
neutralino halo annihilation rates to gamma rays and 
anti-matter are also enhanced relative to 
related scenarios with unified gaugino masses but, depending
on the halo dark matter distribution, may yet be below the
sensitivity of indirect searches underway. 
In the case of collider signals, we compare the rates for the
potentially dominant decay modes of the $\tst_1$ which may be
expected to be produced in cascade decay chains at the LHC: 
$\tst_1\to c\tz_1$ and $\tst_1\to bW\tz_1$. We examine
the extent to which multilepton signal rates are reduced when
the two-body decay mode dominates. For the model
lines that we examine here, the multi-lepton signals, though reduced, 
still remain observable at the LHC.}


\begin{document}

\section{Introduction}
\label{sec:intro}
 
Models of particle physics with weak scale softly broken supersymmetry
are well-motivated
by both theory and experiment. On the theory side, they stabilize
the scalar sector that plays an essential role in the spontaneous
breaking of electroweak symmetry,
%effect
%the cancellation of quadratic divergences which are known to plague
%quantum field theories containing fundamental scalar fields, thus 
allowing a sensible extrapolation of particle interactions over many
orders of magnitude in energy. On the experiment side,
supersymmetric models naturally accommodate, {\it i})~gauge coupling
unification, {\it ii})~a mechanism for electroweak symmetry breaking due
to a large top quark mass, {\it iii})~a light Higgs scalar and decoupled
superpartners in accord with precision electroweak measurements and {\it
iv})~a neutral weakly interacting particle that can, as a thermal Big Bang
relic, account for the observed cold dark
matter (CDM) in the Universe.

In spite of the accolades, supersymmetric theories suffer from new
problems not present in the Standard Model (SM). There are the big
issues such as the flavor and the $CP$ problems, as well as the fact
that baryon and lepton numbers can potentially be violated at large
rates.  We have nothing new to say about these, and will evade these in
the usual ways. A much less serious objection is the ``supersymmetric
little hierarchy problem'' which simply states that the value of the
parameter $-m_{H_u}^2$ (renormalized at the TeV scale) can be $\sim
M_Z^2$ only if there are cancellations at the percent level, once
experimental constraints on sparticle and MSSM Higgs scalar masses are
incorporated.  
%It originates in the translation of the bound
%$m_{\phi}>114.4$ GeV for a SM Higgs boson to a bound on the lightest
%neutral scalar h of the MSSM. For $m_A \agt 200$~GeV, this suggests that
%there must be large radiative corrections to $m_h$ which in turn implies
%that top squarks should be close to the TeV scale.  But heavy top
%squarks naturally imply (via the RGEs) Then, the TeV scale scalars must
%conspire somehow to give a $Z$ mass of $91.17$ GeV (which would require
%some degree of fine-tuning).  
Another potential problem is that in many
supersymmetric models, the lightest SUSY particle, usually the lightest
neutralino, is bino-like, with a typical {\it thermal} relic density 
considerably larger than the measured CDM density
$\Omega_{CDM} h^2\sim 0.1$ \cite{wmap} for sparticle masses larger than
$\sim 100$~GeV.

Recently, Martin has observed that the latter two issues are ameliorated
in a scenario \cite{stevem} that he calls ``compressed
supersymmetry''. Within this framework, it is assumed that the MSSM is
the effective field theory between  $M_{\rm weak}$ and $M_{\rm GUT}$. As
in the mSUGRA model, universal scalar mass parameters are adopted
at $Q=M_{\rm GUT}$
but non-universal gaugino mass parameters are allowed. Specifically,
Martin notes that if $3M_3({\rm GUT})\sim M_2({\rm GUT}) \sim M_1({\rm GUT})$, the
fine-tuning required to obtain small values of $|m_{H_u}^2|$ is considerably
reduced. 
The low value of
$M_3$ results in a SUSY spectrum where physical squark and gluino masses are
closer in mass to uncolored sparticles than in models such
as mSUGRA with unified gaugino masses, where one expects $m_{\tq}\sim
m_{\tg}\gg m_{\tw_1}$.  Thus the SUSY spectrum is ``compressed''
relative to models with gaugino mass unification.

%A compressed spectrum means in particular that there exist solutions
%where the top squark $\tst_1$ can be quite light. 
Of particular interest to us are solutions with a compressed spectrum
where the top squark $\tst_1$ is particularly light.  In this case, if
the neutralino annihilation channel $\tz_1\tz_1\to t\bar{t}$ is
kinematically accessible in the early Universe, its reaction rate
suffers no propagator suppression because of the light $t-$ and $u-$
channel stop exchange, and can lead to a neutralino relic abundance in
accord with WMAP, even though the neutralino remains largely
bino-like. In addition, as noted above, the low third generation squark
masses feed into the evolution of the soft SUSY breaking Higgs mass
$m_{H_u}^2$, causing it to evolve to much smaller (in magnitude)
negative values than in the case of unified gaugino masses. Since
$-m_{H_u}^2(\rm weak) \sim \mu^2$ the little hierarchy problem is less
severe than in models with unified gaugino masses.

Martin has shown that the compressed SUSY scenario is valid provided that
\bea
m_t<m_{\tz_1}\alt m_t+100\ {\rm GeV}, \label{eq1}\\
m_{\tz_1}+25\ {\rm GeV}\alt m_{\tst_1}\alt m_{\tz_1}+100\ {\rm GeV},
\eea
where the lower limits above are imposed so that annihilation
of neutralinos into top pairs is allowed at rest, and to reduce the
impact of $\tst_1$-$\tz_1$ co-annihilation, while the upper limits
should be viewed as soft.  He displays an explicit case where the GUT
scale gaugino masses are related according to
\be
1.5 M_1=M_2 =3M_3,
\ee
which can occur in models where the SUSY breaking $F$-term that seeds SUSY
breaking gaugino masses transforms as a linear combination of 
a singlet and an adjoint field of the unifying $SU(5)$ group.
%the symmetric product of two adjoints, 
The trilinear
soft SUSY breaking term $A_0$ is set either to $-M_1$ or $-0.75 M_1$.
Since the $\tst_1-\tz_1$ mass gap is small in compressed SUSY, Martin 
recognized that two cases
emerge which are relevant to LHC searches: one is characterized by
when $\tst_1\to c\tz_1$ is the dominant top squark decay channel, while
the other has a large enough mass gap that $\tst_1\to bW\tz_1$ can compete,
and perhaps dominate, the two-body decay.

In fact, this whole scenario appears closely related to scenarios first
pointed out by Belanger {\it et al.}\cite{belanger} and independently by
Mambrini and Nezri\cite{mn} and subsequently examined in detail in
Ref. \cite{m3dm}, where a reduced GUT scale gaugino mass $M_3$ leads to
a small $\mu$ parameter, and ultimately to a mixed higgsino-bino $\tz_1$
which can annihilate efficiently into vector boson pairs, ameliorating
the SUSY little hierarchy problem, while in accord with the measured
abundance of cold dark matter in the Universe. While the analyses of
\cite{belanger,mn} and \cite{m3dm} take low $M_3$ in an {\it ad hoc}
fashion, the required gaugino mass pattern can also be obtained by
allowing the SUSY breaking $F$-term to transform as appropriate linear
combinations of fields contained in the symmetric product of two
adjoints of the unifying gauge group\cite{anderson}.  
We note here that a top-down
scenario that naturally leads to low $M_3$, low $|\mu |$ and light top
squarks occurs in so-called mixed moduli-anomaly mediated SUSY breaking
models, also referred to as mirage unification models, wherein moduli
contributions give universal gaugino mass terms, but comparable
gaugino mass splittings from anomaly-mediation reduce the value of
$M_3$, owing to the negative $SU(3)$ beta 
function\cite{mmamsb}\footnote{For a further model with compressed spectra, 
see Bae {\it et al.}, Ref. \cite{bae}.}.

In this paper, we explore the phenomenological implications of
compressed SUSY. We divide our discussion into two different model
lines. In Case A (examined in Sec. \ref{sec:caseA}), we adopt a model
line from Ref.~\cite{m3dm} which is continuously connected to mSUGRA via
variation of the gaugino mass $M_3$, but with a non-zero $A_0$
parameter.  By dialing $M_3$ to smaller values, the top squark mass is
decreased, and the relic density is ultimately dominated by annihilation
to $t\bar{t}$ via light $\tst_1$ exchange. The neutralino, however,
remains essentially bino-like.\footnote{If $M_3$ is reduced farther,
  the neutralino develops a significant higgsino component and leads to
  mixed higgsino dark matter as already mentioned, unless of course,
  this range of $M_3$ is forbidden because $\tst_1$ becomes the LSP.}
The enhanced neutralino
annihilation rate in turn implies an enhanced DM annihilation rate in
the galactic halo\cite{bo}, and we show that indirect DM search rates are thus
enhanced relative to mSUGRA. In addition, the low $\mu$ value and low
$m_{\tq}$ values typical of compressed SUSY result in enhanced rates for
direct DM detection, and detection via muon telescopes.  For this case,
when the measured abundance of CDM is achieved, we arrive at a small
mass gap solution where $\tg\to t\tst_1$ dominantly, followed by
$\tst_1\to c\tz_1$. In addition, the dominant decays $\tw_1\to b\tst_1$
and $\tz_2\to\tz_1 h$ suggest that compressed SUSY LHC signatures are
expected to be {\it lepton poor}, although robust rates for multi-jet
$+\eslt$ signals remain. We note, however, that $\tz_2 \to Z\tz_1$ has
a branching fraction of a few percent. This, combined with the enormous
rate for the production of sub-TeV scale gluinos (in the
dark-matter-allowed regions) makes the multi-lepton signal observable in
the cases we examined. 

In Case B (examined in Sec. \ref{sec:caseB}), we consider a model line
from Martin\cite{stevem} with $1.5 M_1=M_2= 3M_3$. In this case as well,
DM direct and indirect detection rates are larger than for the case of
unified gaugino masses (with large $|\mu|$), and may possibly be
detectable via ton size noble element detectors, or perhaps via
anti-particle and gamma ray searches if the (currently undetermined)
halo dark matter distribution turns out to be suitably clumpy, even
though $\tz_1$ remains dominantly bino-like. Since the
mass gap $m_{\tst_1}-m_{\tz_1}$ can be greater than $ m_b+M_W$, we
implement the 3-body decay $\tst_1\to bW\tz_1$ into Isajet 7.76 (which
we use for spectra and event generation). 
%We also upgrade our treatment
%of $\tst_1\to c\tz_1$ decays, so the relative decay rates can be
%compared.  
We find regions with a large branching fraction for $\tst_1\to bW\tz_1$
decays, so that when this mode dominates, leptonic signals from gluino
and squark cascade decays occur at observable levels.

\section{Case A: Low $M_3$ scenario with continuous connection to mSUGRA}
\label{sec:caseA}

In this section, we examine a model line based on mSUGRA, but with 
$M_3({\rm GUT})$ as an independent parameter, with parameter space
\be
m_0,\ m_{1/2},\ M_3,\ A_0,\ \tan\beta,\ sign(\mu ),
\ee
where we take the GUT scale values\footnote{We will henceforth not
  explicitly specify the scale of the gaugino mass parameters, but this
  should be clear from the context whether we are referring to the
  parameters at the weak or at the GUT scale.} $M_1=M_2\equiv m_{1/2}$
  and adopt $m_t=175$~GeV to conform with Martin\cite{stevem}. The
  phenomenology of this scenario has been investigated in depth in
  Ref. \cite{m3dm} for $A_0=0$, where a low enough value of $M_3 \ll
  m_{1/2}$ leads to a small $\mu$ parameter, and hence the correct dark
  matter relic abundance via mixed higgsino DM. In the case studied
  here, we adopt a value of $A_0=-1.5 m_{1/2}$, which helps reduce
  $m_{\tst_1}$ compared with a choice of $A_0=0$, so that we can obtain
  dominant $\tz_1\tz_1$ annihilation into $t\bar{t}$ via a light
  $\tst_1$ exchange. Of course, if $m_{\tst_1}-m_{\tz_1}$ becomes small
  enough, $\tst_1$-$\tz_1$ co-annihilation will also be important. Since
  for a bino-like LSP $m_{\tz_1}\sim 0.4 m_{1/2}$, we will need
  $m_{1/2}\agt 450$ GeV so that $m_{\tz_1}>m_t$. Thus, we adopt
  $m_{1/2}=500$ GeV, and take $m_0=340$ GeV, $\tan\beta =10$ and $\mu
  >0$ in accord with Martin\cite{stevem}.

The mass spectrum -- generated using Isajet 7.76\cite{isajet} -- is
shown versus $M_3$ in Fig. \ref{fig:Amass}{\it a}). In our illustration,
$M_3=500$ GeV corresponds to the mSUGRA model. Here, the spectrum shows
the well-known feature that the colored sparticles (squarks and gluinos)
are split from, and much heavier than, the lighter uncolored sparticles.
As $M_3$ decreases from 500 GeV, the gluino, and via RGE effects also
squark, masses drop giving rise to the ``compressed SUSY'' mass
spectrum. The $\tst_1$ squark is the lightest of the squarks, owing to
Yukawa coupling and intra-generational mixing effects, and its mass
drops below $m_{\tz_2}$ and $m_{\ttau_1}$ around $M_3\sim 300$ GeV.  We
note that the diminished squark masses feed into the Higgs soft masses
via the RGEs, and give rise to a falling $\mu$ parameter as $M_3$ drops.
The end of parameter space occurs at $M_3\sim 238$ GeV, where the
$\tst_1$ becomes the LSP, and so is excluded by limits on stable charged
or colored relics from the Big Bang. We see that not only $m_{\tz_1}$,
but also $m_{\tz_2}$, is
significantly smaller than $\mu$ even at the lower end of $M_3$ where
the WMAP constraint is satisfied, so though $\tz_1$ develops a
significantly larger higgsino component compared to mSUGRA where it retains
its bino-like character.

In Fig. \ref{fig:Amass}{\it b}), we show the neutralino relic density
$\Omega_{\tz_1}h^2$ versus $M_3$ for the same parameters as in frame {\it a}),
using the IsaReD program\cite{isared}.
For the mSUGRA case of $M_3=500$ GeV, $\Omega_{\tz_1}h^2\sim 1.5$, 
so that the model would be cosmologically excluded, at least if we
assume thermal relics and standard Big Bang cosmology.
As $M_3$ decreases from 500 GeV, $\Omega_{\tz_1}h^2$ drops slowly
until below $M_3\sim 300$ GeV a more rapid fall-off brings 
$\Omega_{\tz_1} h^2$ into accord with the WMAP measurement, which occurs
for $M_3\sim 255$ GeV. At this point, the $\tst_1$ is rather light,
with $m_{\tz_1}\sim 200$ GeV, and $m_{\tst_1}\sim 230$ GeV.
\FIGURE[htb]{ 
\epsfig{file=mass.eps,height=10cm,angle=0}\vspace{2cm} 
\epsfig{file=rd.eps,height=9cm,angle=0} 
\caption{\label{fig:Amass}
{\it a})~Sparticle mass spectrum for the case with 
$m_0=340$~GeV, $M_1=M_2=500$ GeV, $A_0=-1.5 m_{1/2}$, $\tan\beta =10$, 
$\mu >0$ and $m_t=175$ GeV, versus GUT scale $SU(3)$ gaugino mass
parameter $M_3$, and 
{\it b})~neutralino relic density versus $M_3$ for same parameters 
as in frame {\it a}).}}

In Fig. \ref{fig:Asigv}, we show the integrated thermally weighted
neutralino annihilation cross sections times relative velocity versus
$M_3$ as obtained using IsaReD, for various neutralino
annihilation and co-annihilation processes.  Here, $x$ is the
temperature in units of the LSP mass. The neutralino relic density is
determined by the inverse of the sum shown by the solid red line,
%The inverse of these quantities is
%proportional to the neutralino relic density, 
so that large annihilation cross sections yield low relic
densities. In the mSUGRA case with $M_3=500$ GeV, the neutralino
annihilation rate is dominated by annihilation to leptons via
$t$-channel slepton exchange.  As $M_3$ decreases, the squark masses,
and especially the $\tst_1$ mass, decrease, so that $\tz_1\tz_1\to
t\bar{t}$ becomes increasingly important, and in fact dominates the
annihilation rate for $240\ {\rm GeV}<M_3<340$ GeV.  For lower $M_3$
values, the $\tst_1 -\tz_1$ mass gap is below 30 GeV, and top-squark
co-annihilation then dominates, although in this narrow range
$\Omega_{\tz_1}h^2$ does not saturate the measured CDM relic density. We
also see that as $M_3$ decreases, annihilation to $WW$, $ZZ$ and $hh$
also increase in strength due to the lower $\mu$ value, and increasing
higgsino component of the neutralino\footnote{We have traced the 
turnover at low $M_3$ in the various curves to
a drop in the freeze out temperature that determines the range of
integration.}. However, these channels never
dominate in this case.
\FIGURE[htb]{
\epsfig{file=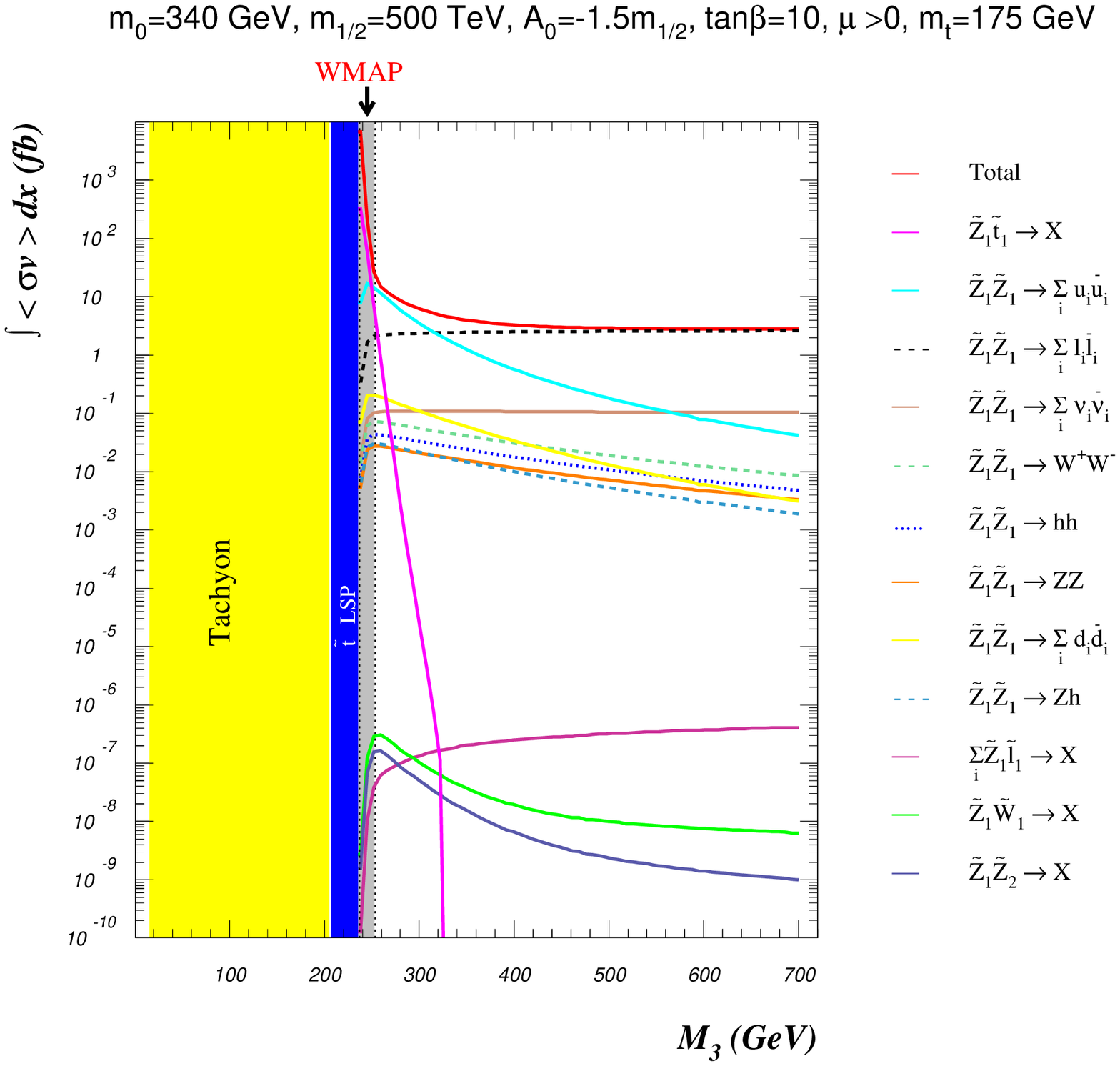,height=10cm} 
\caption{\label{fig:Asigv}
Integrated thermally weighted cross sections times relative velocity 
for processes that may be
relevant for the calculation of the $\tz_1$ relic density in the Big
Bang versus $M_3$. 
%neutralino annihilation (or co-annihilation) cross  sections times
%relative velocity, 
We illustrate these for the same parameters as in 
Fig. \ref{fig:Amass}.
}}

In compressed SUSY, a light top squark is desirable in that it enhances
the neutralino annihilation rate, and brings the relic density
prediction into accord with observation, providing yet another mechanism
for reconciliation of the predicted DM relic density with observation.
However, generically a light top squark also enhances SUSY loop
contributions to the decay $b\to s\gamma$\cite{bb}.  In
Fig. \ref{fig:Absg}, we show the branching fraction $BF(b\to s\gamma )$
vs. $M_3$ for the same parameters as in Fig. \ref{fig:Amass}.  In the
mSUGRA case, the predicted branching fraction is in accord with the
measured value: $BF(b\to s\gamma )=(3.55\pm 0.26)\times 10^{-4}$ from a
combination of CLEO, Belle and BABAR data\cite{bsg_ex}. However, the
light $\tst_1$ in the low $M_3$ region reduces the branching fraction
well below the measured value. Of course, this branching fraction is
also sensitive to other model parameters, {\it e.g.} $\tan\beta$.  The
point, however, is that for the light $\tst_1$ case, the SUSY
contribution is generically comparable to the SM contribution, so that
these must fortuitously combine to be consistent with the experimental
value, which itself is in good agreement with the SM prediction.  At the
very least, in the absence of any real theory of flavor, (such
fortuitous) agreement with the measured value, which agrees well with
the SM prediction \cite{bsg_th} $BF(b\to s\gamma)=(3.29\pm 0.33)\times
10^{-4}$, can always be obtained by allowing a small flavor violation in
the soft parameter matrices at the GUT scale.
% to compensate, but then it would be
%a remarkable co-incidence that the measured value of $BF(b\to s\gamma )$
%agrees so well with the SM prediction\cite{bsg_th}: $BF(b\to s\gamma
%$)=(3.29\pm 0.33)\times 10^{-4}$.
%
\FIGURE[htb]{
\epsfig{file=bsg.eps,height=8cm,angle=0} 
\caption{\label{fig:Absg}
Branching fraction for $b\to s\gamma$ decay  versus $M_3$
for same parameters as in Fig. \ref{fig:Amass}.
}}

\subsection{Scenario A: dark matter searches}
\label{ssec:dm_a}

Next, we investigate prospects for dark matter searches for the case A
model line. We first calculate the spin-independent neutralino-proton
scattering cross section using IsaReS\cite{isares}, and plot the results in
Fig. \ref{fig:Add_mu}{\it a}). In the case of mSUGRA at $M_3=500$ GeV,
the cross section $\sigma_{SI}(\tz_1 p)\sim 10^{-10}$ pb, which is near
the projected limit of future ton-scale noble liquid dark matter
detectors. As $M_3$ decreases, the squark masses also decrease, which
increases the neutralino-proton scattering rate, which occurs primarily
via squark exchange diagrams. Furthermore, a reduced value of $|\mu|$ is
obtained for the low value of $|M_3|$, resulting in an increased
higgsino component of $\tz_1$ (which still remains bino-like) so that
the contribution to the direct detection cross section via the Higgs
exchange diagram is correspondingly increased.  By the time we reach
$\Omega_{\tz_1}h^2\sim 0.1$ at $M_3\sim 255$ GeV, the direct detection
cross section has grown by an order of magnitude, to just above
$10^{-9}$ pb.  This is a general feature of models with a low $M_3$
value\cite{m3dm}: for a given (bino-like) neutralino mass, direct
detection rates are enhanced in the low $M_3$ case.
% due to the compressed spectrum, and lighter squark 
%masses
%\footnote{In models with low $|M_3|$, direct detection rates are also 
%enhanced by Higgs exchange diagrams when the Higgsino component of $\tz_1$ 
%grows due to a small $|\mu |$ parameter.}

In Fig. \ref{fig:Add_mu}{\it b}), we show the flux of muons expected to
be measured at a neutrino telescope from neutralino annihilation into
muon neutrinos in the core of the sun. In this and other indirect
detection rates, we have implemented the Isajet/DarkSUSY
interface\cite{darksusy}. We require muons to have energy $E_\mu >50$
GeV, the threshold for the IceCube detector\cite{icecube}. In this case,
the rate is again enhanced in going from mSUGRA to compressed
SUSY, primarily because of the
diminution of squark mass and the reduced value of $|\mu|$ as already
discussed above: these increase the spin-dependent
neutralino-nucleon scattering cross section and enhance the IceCube rate
because of the increased capture of neutralinos by the sun. 
%The
%second effect is that for low $M_3$, since the dominant annihilation
%channel is to top quark pairs, the neutrino energy spectrum is much
%harder than the mSUGRA case, where the dominant annihilation channel is
%to leptons. {\bf Must check if this is true; it is opposite to case B.}
%Thus, not only the flux increases, but the energy spectrum becomes much
%harder so that the efficiency of muon detection is greatly increased.
In the WMAP-allowed region, the neutralinos mainly annihilate to
$t\bar{t}$ pairs, so that the energy of the neutrino from top decays is
shared with the accompanying $b$ and the daughter muon. We see that 
although the flux of muon neutrinos corresponding to
$E_\mu >50$ GeV increases by a factor of $\sim 500$
in going from mSUGRA to the compressed SUSY case illustrated here, 
the flux of  muon neutrinos is still below the
reach of IceCube, primarily because the neutralino is still mostly bino-like.

\FIGURE[htb]{
\epsfig{file=dd.eps,height=5.2cm,angle=0}\hspace{0.3cm} 
\epsfig{file=idd_mu.eps,height=5.2cm,angle=0} 
\caption{\label{fig:Add_mu}
{\it a}) Spin-independent neutralino-proton scattering cross section 
and {\it b}) flux of muons with $E_\mu >50$ GeV at IceCube 
versus $M_3$ for same parameters as in Fig. \ref{fig:Amass}.
}}

For positrons and anti-protons, 
we evaluate the averaged differential antiparticle flux in
a projected energy bin centered at a kinetic energy of 20~GeV, where we
expect optimal statistics and signal-to-background ratio at
space-borne antiparticle detectors\cite{statistical}. We take
the experimental sensitivity to be that of the Pamela experiment after
three years of data-taking as our benchmark\cite{pamela}.
The expected fluxes depend on the (unknown) details of the
neutralino distribution in our galactic halo.
Here, we assume
a scenario where baryon infall causes progressive deepening of the
gravitational potential well, and a clumpy halo distribution
is obtained: the Adiabatically Contracted $N03$ Halo Model\cite{achm}

In Fig. \ref{fig:A_antim} we show the expected positron flux in frame
{\it a}) and the expected anti-proton flux in frame {\it b}) versus
$M_3$ for the same parameters as in Fig. \ref{fig:Amass}. We see that in
each case the antimatter flux jumps by a factor of $\sim 10^2$ in going
from mSUGRA to compressed SUSY, largely due to the enhanced annihilation
rate into $t\bar{t}$ pairs, and the concomitant hard spectrum of $e^+$s
and $\bar{p}$s that ensue. In the case shown, the positron flux for
compressed SUSY is somewhat below the Pamela reach, while the $\bar{p}$
flux is near the Pamela reach. We warn the reader that for the smooth
Burkert halo profile \cite{burkert} the signals are significantly
smaller and beyond the sensitivity of Pamela.
\FIGURE[htb]{
\epsfig{file=idd_ep.eps,height=5.2cm,angle=0} \hspace{0.3cm} 
\epsfig{file=idd_pb.eps,height=5.2cm,angle=0} 
\caption{\label{fig:A_antim}
{\it a}) Expected positron flux 
and {\it b}) antiproton flux 
versus $M_3$ for same parameters as in Fig. \ref{fig:Amass}. The dashed
line shows the expected three year sensitivity of Pamela.
}}

We have also evaluated the average differential anti-deuteron flux in
the $0.1<T_{\bar D}<0.25$ GeV range, where $T_{\bar D}$ stands for the
antideuteron kinetic energy per nucleon, and compared it to the estimated
sensitivity of GAPS for an ultra-long duration balloon-borne
experiment\cite{gaps}.  We see in Fig. \ref{fig:A_dbgam}{\it a}) that
the antideuteron flux is again enhanced by a factor of $\sim 10^2$ in going
from mSUGRA to compressed SUSY, and in fact moves above the
detectability limit of the GAPS experiment. For the Burkert halo
profile, the estimated flux for the WMAP-allowed range of $M_3$ is
essentially at the edge of detectability. 

Indirect detection of neutralinos is also possible via the detection of
high energy gamma rays\cite{egret} produced by neutralino
annihilation in the center of our Galaxy\cite{gammas}. These will also
be searched for by the GLAST collaboration \cite{glast}.  We have evaluated
expectations for the integrated continuum $\gamma$ ray flux above an
$E_\gamma=1$ GeV threshold versus $M_3$ in Fig. \ref{fig:A_dbgam}{\it b}). 
These projections are
extremely sensitive to the assumed neutralino halo distribution, and
drop by more than four orders of magnitude for the Burkert halo profile.
This makes it difficult to make any definitive statement about the
detectability of this signal (which could serve to map the
halo profile rather than a diagnostic of the nature of the DM particle).
However, once again we see a factor of $\sim 100$ enhancement in detection 
rate in moving from the mSUGRA case where $M_3=500$ GeV to the compressed 
SUSY case with $M_3\sim 255$ GeV.
\FIGURE[htb]{
\epsfig{file=idd_db.eps,height=5.2cm,angle=0} \hspace{0.3cm}
\epsfig{file=idd_gam.eps,height=5.2cm,angle=0} 
\caption{\label{fig:A_dbgam}
{\it a}) Expected anti-deuteron flux 
and {\it b}) gamma ray flux 
versus $M_3$ for same parameters as in Fig. \ref{fig:Amass}. The
horizontal lines show the projected sensitivities of the GAPS and GLAST
experiments. 
}}

\subsection{Scenario A: LHC searches}
\label{ssec:lhc_a}

As Martin notes\cite{stevem}, the compressed SUSY mass spectra are
generally too heavy for successful sparticle searches at the Fermilab
Tevatron. However, (\ref{eq1}) implies an upper bound on the bino mass,
and since we {\it reduce} $M_3$ from its unified value, implies that
gluinos must be relatively light so that
 multi-jet $+$ multilepton $+\eslt$ events from SUSY
should be produced in abundance at the CERN LHC, due to turn on in
2008. In this section, we investigate the collider signals expected
after cuts for various signal topologies at the LHC.

At the CERN LHC, gluino and squark pair production will be the dominant
SUSY production reactions. Gluino and squark production will be followed
by their cascade decays\cite{bbkt}, resulting in a variety of events
with jets, isolated
leptons and missing energy.  A large number of signals
emerge, and can be classified by the number of isolated leptons
present. The signal channels we examine include {\it i.})~no isolated
leptons plus jets plus $\eslt$ ($0\ell$), {\it ii.})~single isolated
lepton plus jets plus $\eslt$ ($1\ell$), {\it iii.})~two opposite sign
isolated leptons plus jets plus $\eslt$ (OS), {\it iv.})~two same sign
isolated leptons plus jets plus $\eslt$ (SS) and {\it v.})~three
isolated leptons plus jets plus $\eslt$ ($3\ell$).

The reach of the CERN LHC for SUSY has been estimated for the mSUGRA
model in Ref. \cite{bcpt1,cms} for low values of $\tan\beta$ and in
Ref. \cite{lhcltanb} for large $\tan\beta$ values.  We adopt the cuts and
background levels presented in Ref. \cite{bcpt1} for our analysis of the
signal channels listed above.  Hadronic clusters with $E_T>100$ GeV and
$|\eta ({\rm jet})|<3$ within a cone of size $R=\sqrt{\Delta\eta^2
+\Delta\phi^2} =0.7$ are classified as jets.  Muons and electrons are
classified as isolated if they have $p_T>10$ GeV, $|\eta (\ell )|<2.5$,
and the visible activity within a cone of $R =0.3$ about the lepton
direction is less than $E_T({\rm cone})=5$ GeV.

Following Ref. \cite{bcpt1}, we required that the jet multiplicity,
$n_{\rm jet} \geq 2$, transverse sphericity $S_T > 0.2$, $E_T(j_1)$, and
further, that $E_T(j_2) \ > \ E_T^c$ and $\eslt > E_T^c$, where the cut
parameter $E_T^c$ is chosen to roughly optimize the signal from gluino
and squark production.  For the leptons we require $p_T(\ell) > 20$~GeV
($\ell=e$ or $\mu$) and $M_T(\ell,\eslt) > 100$~GeV for the $1\ell$
signal. For the $OS$, $SS$ and $3\ell$ channels, we require that the two
hardest leptons have $p_T \ge 20$~GeV.  We have also applied a cut on
the transverse plane angle $\Delta \phi(\vec{E}_T^{\rm miss},j_c )$
between $\vec{E}_T^{\rm miss}$ and closest jet: $30^\circ<\Delta\phi
<90^\circ$, in the case of the $\eslt$ channel, $i)$.

Our results are shown in Fig. \ref{fig:Alhc} for a rather loose 
choice of the cut
parameter $E_T^c =100$~GeV.
%The solid horizontal mark on each curve denotes the minimum cross section
%needed for discovery incorporating three criteria:
%{\it i})~the signal to background ratio, $S/B \ge 0.2$, {\it ii})~the signal has a minimum of
%five events, and {\it iii})~the signal satifies a statistical criterion
%$S \ge 5\sqrt{B}$ for an integrated luminosity of 100~fb$^{-1}$.
We see that as $M_3$ decreases from the mSUGRA value of $500$ GeV, the
signal cross sections increase. The increase is mainly due to increased
total gluino and squark production cross sections, due to their
decreasing masses. When we reach the DM -allowed compressed SUSY
spectrum at $M_3\sim 250$ GeV, however, the leptonic signals suffer a
steep drop-off, while the $\eslt +$ jets signal increases somewhat. This
is due to the fact that in this case, $\tw_1\to b\tst_1$ turns on and
dominates the $\tw_1$ branching fraction, while $\tst_1\to c\tz_1$ at
essentially 100\%. Thus, no isolated leptons come from chargino decay.
Likewise, $\tz_2\to \tz_1h$ at around 90\% branching fraction, so
isolated leptons from $\tz_2$ decays come from the subdominant decay
chain $\tz_2 \to \tz_1 Z$ which has a branching fraction of $\sim 8$\%.
Isolated leptons still arise from $\tg\to t\tst_1$ decay, followed by
semi-leptonic top decay, but in general, we expect in compressed SUSY
models with a small $\tst_1 -\tz_1$ mass gap and $m_{\tw_1}>m_{\tst_1}
+m_b$ that the fraction of signal events containing isolated leptons
will be much {\it lower} than the usual prediction from models like
mSUGRA with gaugino mass unification.  We regard a signal to be
observable if for the given integrated luminosity, {\it i})~the
statistical significance of the signal exceeds $5\sigma$, {\it ii})~$S/B
> 0.25$, and {\it iii})~$S > 10$~events. The minimum observable cross
sections for each topology are shown by the dashed horizontal bars in
the figure. We see that even for the low value of $E_T^c=100$~GeV, all
but the opposite sign dilepton signal should be observable with an
integrated luminosity of 100~fb$^{-1}$, and frequently even with a much
lower integrated luminosity, at least for parameters in the WMAP-allowed
region. Although we do not show this, we have checked that with
$E_T^c=200$~GeV, the OS signal is easily observable,\footnote{Since the
OS dileptons come primarily from the decay of an on-shell $Z$ boson, it
is possible that this signal would actually be observable even for
$E_T^c=100$~GeV.}  and furthermore, the $0\ell$ signal is not as close
to the observability limit.

\FIGURE[htb]{
\epsfig{file=lepsig_A.eps,height=12cm,angle=0} 
\caption{\label{fig:Alhc} Signal rates at the CERN LHC for various
multi-jet plus multi-lepton $+\eslt$ event topologies after cuts listed
in the text with the cut parameter $E_T^c=100$~GeV versus $M_3$ for same
parameters as in Fig. \ref{fig:Amass}. The horizontal dotted lines show
the minimum observable cross section for $E_T^c=100$~GeV, assuming an
integrated luminosity of 100~fb$^{-1}$. }}

\section{Case B: Non-universal gaugino masses
and a large mass gap}
\label{sec:caseB}

In this section, we explore Case B, the compressed SUSY model line
originally suggested by Martin where at $Q=M_{\rm GUT}$, $1.5
M_1=M_2=3M_3$, with $m_0=340$ GeV, $A_0=-0.75 M_1$, $\tan\beta =10$ and
$\mu >0$. We first display the variation of the 
sparticle mass spectrum with
$M_1$ in Fig. \ref{fig:Bmass}{\it a}). The upper end of parameter
space is limited by $M_1\alt 1000$ GeV, where for higher $M_1$ values
the $\tst_1$ becomes the LSP. This implies an upper bound of 1200~GeV
(1100-1400~GeV) on gluino (squark) masses, ensuring their copious
production at the LHC. The lower range of $M_1$ is bounded by
$M_1\agt160$ GeV, since for lower $M_1$ values, the value of $m_{\tw_1}$
drops below limits from LEP2 \cite{lep2}. In the intermediate region
with $440\ {\rm GeV}<M_1<1000$ GeV, the $\tst_1$ is relatively light,
and is the next-to-lightest SUSY particle (NLSP). More importantly from
our perspective, in this mass regime $m_{\tz_1}>m_t$, ensuring  that
$\tz_1\tz_1\to t\bar{t}$ was accessible in the early Universe.

In Fig. \ref{fig:Bmass}{\it b}), we show the neutralino relic density as
a function of $M_1$ for the same parameters as in frame {\it a}). There
is a wide range of $M_1: 400-800$ GeV where the relic abundance is in
close accord with the WMAP measured value. It should be possible to
bring this in accord with the WMAP value by slightly tweaking $A_0$.
For yet larger values of $M_1$, $\tst_1\tz_1$ and $\tst_1\tst_1$
annihilation rates become large, and the $\tz_1$ relic density no longer
saturates the observed density of CDM; {\it i.e.} the DM would be
multi-component in this case.  In contrast, when $M_1$ drops below $\sim
400$ GeV, corresponding to $m_{\tz_1}<m_t$, the prediction for
$\Omega_{\tz_1}h^2$ rises above the WMAP measurement, excluding $\tz_1$
as a thermal relic.  For $M_1\sim 150$~GeV -- a range excluded by
the LEP2 chargino mass limit -- there is a double dip structure where
$2 m_{\tz_1}\sim m_h$ or $M_Z$, and so neutralinos can efficiently
annihilate through these $s-$channel poles.

\FIGURE[htb]{\vspace{2cm}
\epsfig{file=mass_b.eps,height=10cm,angle=0}  
\epsfig{file=rd_b.eps,height=9cm,angle=0} 
\caption{\label{fig:Bmass} {\it a}): Sparticle mass spectrum as a
function of the GUT scale gaugino mass $M_1$ for Case~B, where
$m_0=340$, $1.5 M_1=M_2=3M_3$ GeV, $A_0=-0.75 M_1$, $\tan\beta =10$,
$\mu >0$ and $m_t=175$ GeV.  {\it b}): Neutralino relic density versus
$M_1$ for same parameters as in frame {\it a}).}}

In Fig. \ref{fig:Bsigv}, we show the integrated thermally weighted
neutralino annihilation (and co-annihilation) cross section times
relative velocity versus $M_1$ for the same parameters as in
Fig. \ref{fig:Bmass}. For $M_1\agt 750$ GeV, the $\tst_1 -\tz_1$ mass
gap is so low that $\tz_1 \tst_1$ co-annihilation, and eventually
$\tst_1\tst_1$
annihilation (not shown in the figure), dominates and we get too small a
relic abundance. In the range 400~GeV $\alt M_1\alt$ 750~GeV, $\tz_1\tz_1\to
t\bar{t}$ dominates, so agreement with the relic density is obtained as
envisioned by Martin \cite{stevem}. For $M_1\alt 400$ GeV, annihilation
into $t\bar{t}$ is not allowed (except for $\tz_1$s in the high energy
tail of the thermal distribution), and so annihilation takes place
dominantly into $WW$ (via the higgsino component) and into quarks and
leptons. At the $h$ and $Z$ poles (inside the LEP-forbidden region),
annihilation into down-type fermions dominates.
\FIGURE[htb]{\epsfig{file=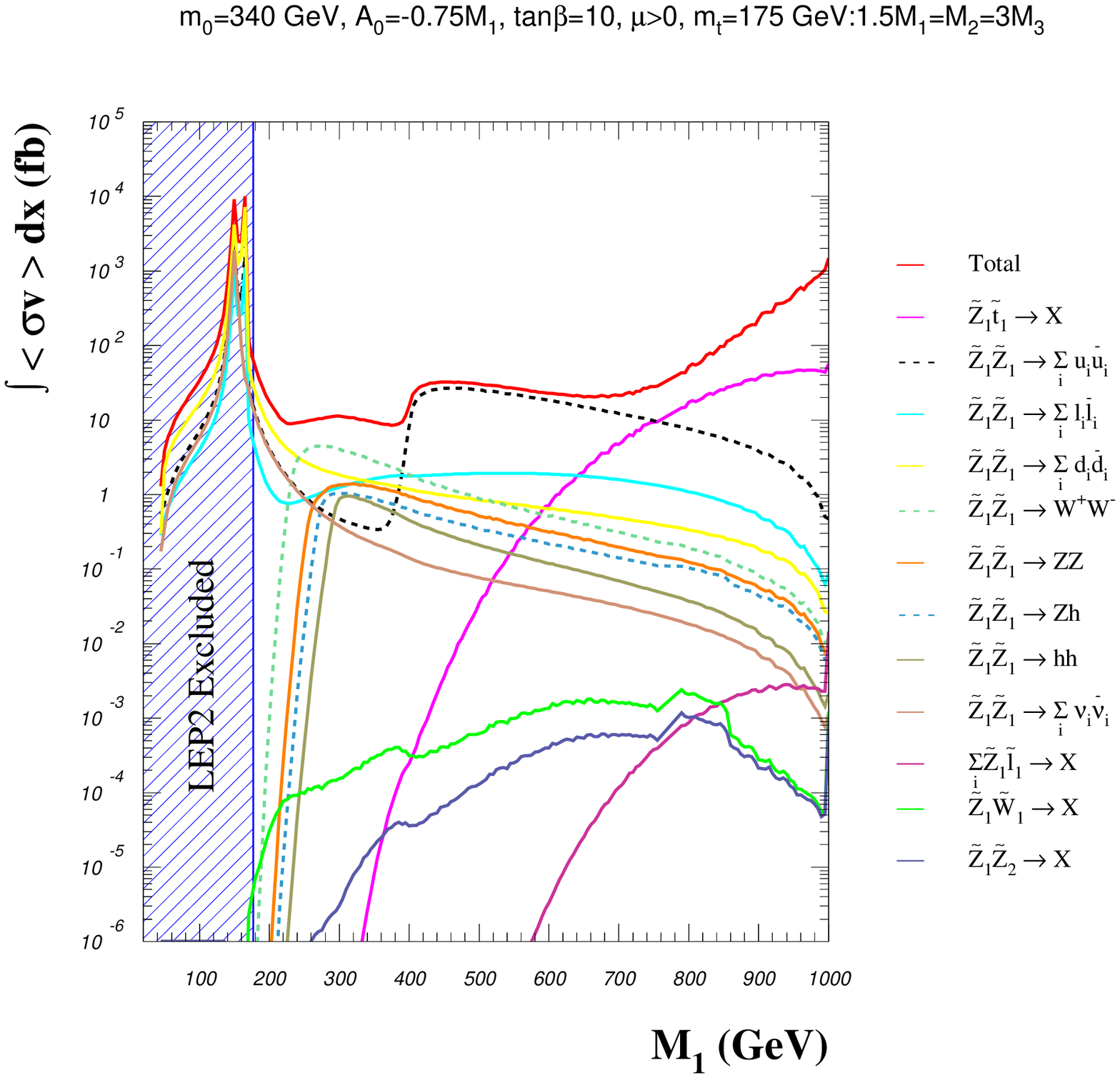,height=10cm}
\caption{\label{fig:Bsigv} Integrated thermally weighted neutralino
annihilation (or co-annihilation) cross sections times relative
velocity, for same parameters as in Fig. \ref{fig:Bmass}, versus
$M_1$. The processes shown do not saturate the total at very large
values of $M_1$ because we have not plotted $\tst_1\tst_1$ annihilation
which becomes very important there because $\tst_1$ becomes very close to
$m_{\tz_1}$ at the upper end of the $M_1$ range. }}

The branching fraction $BF(b\to s\gamma )$ is shown in
Fig. \ref{fig:Bbsg} versus $M_1$ for the same parameters as in
Fig. \ref{fig:Bmass}. Here we note that for large $M_1$, the branching
fraction is close to-- albeit somewhat below-- its measured  value.
However, as
$M_1$ decreases, the $\tst_1$ and $\tw_1$ both become lighter, and
SUSY loop contributions to the branching fraction move the 
predicted branching
fraction away from its observed value. In
this case, as in Sec. \ref{sec:caseA}, we would expect a somewhat
suppressed value of $BF(b\to s\gamma )$ compared to its SM predicted
rate. We recall as before that it should be possible to bring
this result into accord with experiment by allowing, for instance, some
flavor structure in the soft SUSY breaking sector.
\FIGURE[htb]{\vspace{0.2cm}
\epsfig{file=bsg_b.eps,height=7.5cm,angle=0} 
\caption{\label{fig:Bbsg} Branching fraction for $b\to s\gamma$ decay
versus $M_1$ for Case~B, for the same model parameters as in
Fig. \ref{fig:Bmass}.  }}

\subsection{Case B: dark matter searches}
\label{ssec:dm_b}

Here, we examine direct and indirect dark matter detection rates for the
compressed SUSY model line Case B. We begin by considering the prospects
for direct detection in Fig. \ref{fig:Bdd_mu}{\it a}) where we show the
spin-independent $\tz_1 p$ cross section as a function of the GUT scale
parameter $M_1$. The cross section increases as $M_1$ decreases due to
decreasing squark masses, and a decreasing value of the $\mu$ parameter.
The range relevant for compressed SUSY, $M_1:400-750$ GeV, has
$\sigma_{SI}(\tz_1 p)\sim 5-15\times 10^{-9}$ pb, which at its high
end is within an order of magnitude of the current limit from 
XENON-10\cite{xenon10},
and should be detectable by SuperCDMS or 100-1000 kg noble
liquid DM detectors. Projections for direct detection are somewhat more
optimistic than in Case~A, mostly because the value of $\mu$ is relatively
smaller in Case~B.

In frame {\it b}), we show the flux of muons with $E_\mu>50$
GeV expected at neutrino telescopes due to neutralino annihilation in
the solar core. As $M_1$ decreases from 1000 GeV, the rate slightly
increases, due to an increasing spin-dependent $\tz_1$-nucleon
scattering rate, but  for the $M_1$ range
of interest, the flux remains somewhat below the IceCube detectable level. For
$M_1<400$ GeV, the rate jumps to higher levels. This jump can be
understood from Fig. \ref{fig:Bsigv}, from which we infer that 
since the neutralino capture and annihilation processes are in equilibrium,
the fraction of captured neutralinos that directly annihilate
into $\nu\bar{\nu}$ jumps once annihilation
to $t\bar{t}$ turns off, and it is these very high energy neutrinos
which have the greatest chance of being detected by IceCube. For
$M_1>400$ GeV, $\tz_1\tz_1$ annihilates mainly into $t\bar{t}$, and the
fraction of direct neutralino annihilation into neutrinos is lower.
\FIGURE[htb]{
\epsfig{file=dd_b.eps,height=5.2cm,angle=0}\hspace{0.3cm} 
\epsfig{file=idd_mu_b.eps,height=5.2cm,angle=0} 
\caption{\label{fig:Bdd_mu} {\it a}) Spin-independent neutralino-proton
scattering cross section and {\it b}) flux of muons with $E_\mu >50$ GeV
at IceCube versus $M_1$ for Case~B with the same parameters as in
Fig. \ref{fig:Bmass}.  }}

In Fig. \ref{fig:B_antim} we show the flux of {\it a}) positrons and
{\it b}) anti-protons from neutralino annihilations in the
galactic halo expected in Case B versus $M_1$, assuming the clumpy halo
as given by the adiabatically contracted N03 halo model.  We evaluate
the signal in the same energy bins and apply the same sensitivity
criterion as in Fig. \ref{fig:A_antim}. The flux of $e^+$s is everywhere
below the Pamela sensitivity even for our favorable choice of halo
distribution. However, the results do show some structure and
enhancement in the compressed SUSY range of $M_1: 440-800$ GeV. In this
regime, $m_{\tz_1}>m_t$ so that $\tz_1\tz_1\to t\bar{t}$ can occur in
the present galactic halo as well as in the early Universe. The turn-on
of the $t\bar{t}$ annihilation mode is clearly seen at $M_1\sim 440$
GeV.  In the case of the $\bar{p}$ flux, the signal actually increases
enough to suggest some range of observability at Pamela.
\FIGURE[htb]{
\epsfig{file=idd_ep_b.eps,height=5.2cm,angle=0} \hspace{0.3cm} 
\epsfig{file=idd_pb_b.eps,height=5.2cm,angle=0} 
\caption{\label{fig:B_antim} {\it a}) Expected positron flux and {\it
b}) antiproton flux for Case~B versus $M_1$ for same parameters as in
Fig. \ref{fig:Bmass}.  }}

In Fig. \ref{fig:B_dbgam} we show {\it a}) the flux of anti-deuterons 
along with the reach of the GAPS experiment, and {\it b}) the flux of
gamma rays from the galactic center with $E_\gamma >1$ GeV.
In the case of $\bar{D}$s, the entire compressed SUSY range is
above the GAPS sensitivity. We caution, however, that for the smooth
Burkert halo profile, projections could be lower by a factor 10-15. 
For $\gamma$s, the entire range should be probed by GLAST, although these
projections are extremely sensitive to our assumed halo
distribution; for other halo choices -- such as the Burkert profile, the rates 
scale downwards by over four orders of magnitude, and could fall below
the projected sensitivity of GLAST.
However, in both the case of $\bar{D}$s and $\gamma$s, a sharp increase in
indirect detection rate occurs when $\tz_1\tz_1\to t\bar{t}$ turns on at
$M_1\sim 440$ GeV.

\FIGURE[htb]{
\epsfig{file=idd_db_b.eps,height=5.2cm,angle=0} \hspace{0.3cm} 
\epsfig{file=idd_gam_b.eps,height=5.2cm,angle=0} 
\caption{\label{fig:B_dbgam} {\it a}) Expected anti-deuteron flux and
{\it b}) gamma ray flux for the compressed SUSY Case~B versus $M_1$ for
same parameters as in Fig. \ref{fig:Bmass}.  }}

\subsection{Scenario B: LHC searches}
\label{ssec:lhc_b}

An important issue for evaluating collider signals in compressed SUSY with
a large $\tst_1-\tz_1$ mass gap is to evaluate the $\tst_1$ decay
branching fractions correctly when the $\tst_1$ is the NLSP. In this
case, the flavor changing decay $\tst_1\to c\tz_1$ may compete with the
three-body decay $\tst_1\to b W\tz_1$ if the latter decay mode is
kinematically allowed. We implement the three body decay into Isajet
7.76 using the squared matrix element calculated by Porod and
W\"ohrmann\cite{pw}.\footnote{We have, however, made one correction from
their erratum. The term $2m_{\tz_1}^2\left(2p_b\cdot p_W +
m_{\tz_1}^2\right)$ in Eq.~(A.2) should be replaced by $2m_{\tz_1}^2
p_b\cdot p_W$ and not by $4m_{\tz_1}^2 p_b\cdot p_W$ as stated in their
erratum. M.~M\"uhlleitner (private communication) has independently
confirmed this factor, which also appears correctly in the program
SDECAY\cite{sdecay}.} 
We also update the Isajet formulae for the flavor changing
two-body decay using the one-step integration approximation of Hikasa
and Kobayashi\cite{hk}, but with the correct neutralino eigenvectors
\cite{gunpap}.  We have checked that this single step integration
systematically over-estimates the width of the loop decay $\tst_1 \to
c\tz_1$, so that if we find the three body decay to be dominant within
our approximation, this will be the case also with the correct
calculation. In Fig. \ref{fig:t1bf}, we show the branching fraction of
$\tst_1$ versus $M_1$ for the same parameters as in
Fig. \ref{fig:Bmass}. We see that at large $M_1$ where
$m_{\tst_1}<m_b+M_W+m_{\tz_1}$, the $\tst_1$ decays entirely into
$c\tz_1$\footnote{There are four body decay modes such as $\tst_1\to
bf\bar{f}'\tz_1$ (where $f$ is a SM fermion) which we have not
evaluated, but which we expect to be smaller than the two-body
decay\cite{djouadi}.}.  For lower $M_1$ values, the $\tst_1\to bW\tz_1$
decay mode opens up and in fact dominates the two-body mode for $M_1:
400-460$~GeV.  For $M_1\alt 400$ GeV, then $m_{\tst_1}>m_b +m_{\tw_1}$,
so that $\tst_1\to b\tw_1$ turns on and dominates the branching
fraction. In this regime, for this case at least, $m_{\tz_1}<m_t$, so
this range is not as interesting from the perspective  of obtaining agreement
with the relic density measurement via neutralino annihilation to top
quarks.
%technically we
%are out of the compressed SUSY parameter space as defined by
%Martin\cite{stevem}.
%
\FIGURE[htb]{
\epsfig{file=bft1.eps,height=8cm,angle=0} 
\caption{\label{fig:t1bf}
Branching fraction of $\tst_1$ vs. GUT scale gaugino mass $M_1$
for same parameters as in Fig. \ref{fig:Bmass}.
}}

Once the correct decay patterns are implemented, we can generate
collider events and examine signal rates after cuts.  We present
multi-lepton plus multi-jet $+\eslt$ cross sections at the LHC for Case
B, using the same analysis as in Sec. \ref{ssec:lhc_a}, except with
$E_T^c=200$~GeV. The results are shown in Fig. \ref{fig:Blhc}.  For low
$M_1$ values, the squark and gluino masses are relatively light, and
SUSY particle production cross sections are large at the CERN LHC.
Nevertheless, signals in the $0\ell$, $1\ell$ and $OS$ channels fall
below the $S/B \ge 0.25$ level for $M_1$ in the interesting range of
400-800~GeV if we choose $E_T^c=100$~GeV. In contrast, with
$E_T^c=200$~GeV shown in the figure, signals in all channels are
observable for the entire range of $M_1$.\footnote{The $E_T^c=100$~GeV
is better optimized for the signals in the $SS$ and $3\ell$ channels.}
As $M_1$ increases, the $\tg -\tw_1$ and $\tq -\tw_1$ mass gaps actually
increase, and we get an increase in the multi-lepton signal rates.
These show a rapid drop off beyond $M_1=450$-500~GeV, where $\tst_1$
produced in gluino cascades decay via $\tst_1 \to c\tz_1$. There is no
analogous drop-off in the $0\ell$ or even in the $1\ell$ channels (since
it is not difficult to get a single lepton somewhere in the cascade,
{\it e.g} via the decay of $t$).  Thus, once the stop is light enough so
it can only decay via $\tst_1 \to c\tz_1$ (which is the case over most
of the $M_1$ range of interest), we see a relative reduction of
multi-leptonic signals compared with those containing just $\eslt
+$jets. Distinguishing Case~B (with $\tst_1 \to c\tz_1$) from Case~A
will be challenging at the LHC, but should be straightforward at a TeV 
linear collider. 
\FIGURE[htb]{
\epsfig{file=lepsig_etc200_B.eps,height=10cm,angle=0} 
\caption{\label{fig:Blhc} Signal rates for Case~B for various multi-jet
plus multi-lepton $+\eslt$ events at the CERN LHC, after cuts detailed
in the text and $E_T^c=200$~GeV, versus $M_1$ for same parameters as in
Fig. \ref{fig:Bmass}. The horizontal dotted lines show the minimum
observable cross section for $E_T^c=200$~GeV, assuming an integrated
luminosity of 100~fb$^{-1}$.  }}

\section{Summary and conclusions}\label{sec:conclude}

The {\it generic} prediction of the neutralino relic density 
from SUSY models falls somewhat above the measured value 
if sparticles are significantly heavier than $\sim 100$~GeV, 
as is likely to be the case given the direct constraints from LEP2 
and the Tevatron, and indirect constraints from low energy measurements. 
Within a particular framework
such as mSUGRA, this means special regions of parameter space where at
least one of: neutralino co-annihilation with staus/stops, neutralino
resonance annihilation via $A/H$ (this requires large $\tan\beta$
values) or even $h$, or mixed higgsino DM in the hyperbolic branch/focus
point region at large $m_0$, obtains. Each of these alternatives would
have implications for SUSY signals both at colliders, as well as for
direct and indirect searches for DM. Unfortunately, these implications
are not robust to small changes in the model. Allowing for
non-universality of gaugino or Higgs scalar mass parameters leads to
one-parameter extensions of the mSUGRA model where the implications of
the WMAP measurement can be strikingly different. For instance,
non-universal Higgs mass models allow mixed Higgsino DM for low values
of $m_0$, and Higgs resonance annihilation for all values of $\tan\beta$
\cite{nuhm}. Non-universal gaugino masses allow new possibilities, such
as mixed wino DM \cite{mwdm} or bino-wino co-annihilation \cite{bwca}
that are precluded in models with unified gaugino masses (but realized
in other frameworks). These studies suggest that it would be premature
to blindly use the measured relic density to make definitive projections
for what should/should not be seen at the LHC or in DM searches. Already
there exist  numerous alternatives (with different
phenomenological outcomes) to choose from, and only experiment can zero
in on nature's choice.

In this vein, Martin \cite{stevem} recently pointed out yet another
possibility to obtain agreement with the observed CDM relic density.  He
noted that if $m_{\tz_1}> m_t$ and $\tst_1$ is not much heavier than
$\tz_1$, then $\tz_1\tz_1 \to t\bar{t}$ mediated by $\tst_1$ exchange
(this process does not suffer the large $p$-wave suppression on account
of the large top mass) may dominate in the early universe. This scenario
can be realized for a bino-like $\tz_1$ only if  gluinos (and through
the RGEs, also the squarks)  are not very heavy, leading to a
``compressed SUSY'' spectrum. In this paper, we have examined two
different model lines that realize Martin's idea, and quantified the
implications for SUSY searches at the LHC as well as via direct and
indirect searches for DM.

The first model line that we refer to as Case~A is continuously
connected to mSUGRA, and is in a sense an extension of our earlier work
that we referred to as low $|M_3|$ dark matter, where relaxing the
gaugino mass unification condition and allowing $|M_3({\rm GUT})|$ to be
smaller than $M_1\sim M_2$ led to viable solutions with mixed higgsino DM
\cite{m3dm}. In these studies, we used $A_0=0$ for simplicity. Here, we
choose instead $A_0=-1.5m_{1/2}$, and lower $M_3$ as before. This choice of
$A_0$ leads to a reduction in $m_{\tst_1}$, and remarkably, as $M_3$ is
reduced, $m_{\tst_1}$ becomes close to $m_{\tz_1}$, so that $\tz_1\tz_1
\to t\bar{t}$ can indeed be the dominant mechanism in the early
Universe, with $\tz_1$ retaining its bino-like character. While the
reduced gluino, and concomitantly squark, masses and the $\mu$
parameter, imply larger direct and indirect detection rates {\it vis \`a
vis} models with gaugino mass unification, these rates are not large,
primarily because the neutralino remains bino-like. Nevertheless ton
size noble liquid detectors should be able to directly see a WIMP signal
(at least for parameters that give the observed relic density), while
indirect searches for anti-deuteron at GAPS or gamma rays from our
galactic center by GLAST may yield an observable signal, but only if the
DM is favorably clumped. We project that there will be no detectable signal in
Pamela or in IceCube.  The scenario implies that gluinos and squarks
cannot be too heavy so that the LHC should be awash in SUSY events, and
the signal should be extricable from SM backgrounds with simple
cuts. The characteristic feature of the scenario is the relative
reduction of the signal in multi-lepton channels relative to that in
$0\ell$ or $1\ell$ channels.  The large production rate nevertheless
implies there should be an observable signal in {\it all} the channels
shown in Fig.~\ref{fig:Alhc}.  A significant fraction of $OS$ dilepton
and trilepton events may contain a real $Z$ boson.

The second model line that we examine (and refer to as Case~B) is the
one suggested by Martin in his original proposal. Here, we adopt
non-universal boundary conditions $1.5M_1=M_2=3M_3$ for the GUT scale
gaugino mass parameters. Prospects for direct detection may be somewhat
better in this scenario: in favorable cases, the signal cross section
may be just an order of magnitude away from the current upper
bound. Indirect detection prospects are similar to those in
Case~A. There is no detectable signal at IceCube, potentially observable
signals in GLAST or GAPS for favorable halo distributions, and possibly
a marginal signal from $\bar{p}$ in Pamela. Experiments at the LHC
should be able to detect a signal in all channels, albeit with somewhat
harder cuts than in Case~A, as illustrated in Fig.~\ref{fig:Blhc}. As in
Case~A, over most of the parameter range compatible with the relic
density measurement, multi-lepton signals will occur at smaller rates.

A light $\tst_1$ is the hallmark of the scenario. While its direct
detection is not easy at the LHC,\footnote{The techniques suggested in
Ref.~\cite{nojiri} do not apply since $\tst_1 \to c\tz_1$.} its presence
along with that of a not-too-heavy chargino leads to a significant SUSY
contribution to the $b\to s\gamma$ branching ratio, and likely also to
the branching ratio and distributions for $b\to s\ell\bar{\ell}$ decays
(that we have not examined). Indeed, for both cases that we examined,
the former turns out to be smaller than its measured value. While it is
certainly true that we can always reproduce the observed 
branching fraction by tweaking the flavour
structure of soft-SUSY-breaking parameters, it would seem unlikely this
would be ``just right'' to yield the SM prediction. It, therefore, seems
that a deviation of the patterns of rare flavor-violating decays of
$b$-quarks from SM expectations should generically be expected in these
scenarios.

\acknowledgments

This research was supported in part by the U.S. Department of Energy
grant numbers DE-FG02-97ER41022 and DE-FG02-04ER41291.

% ---- Bibliography ----
%

\end{document}